# Combining docking pose rank and structure with deep learning improves protein-ligand binding mode prediction

Joseph A. Morrone*, Jeffrey K. Weber, Tien Huynh, Heng Luo, Wendy D. Cornell

Healthcare & Life Sciences Research, IBM TJ Watson Research Center

1101 Kitchawan Road, Yorktown Heights, NY 10598, USA

**ABSTRACT:** We present a simple, modular graph-based convolutional neural network that takes structural information from protein-ligand complexes as input to generate models for activity and binding mode prediction. Complex structures are generated by a standard docking procedure and fed into a dual-graph architecture that includes separate sub-networks for the ligand bonded topology and the ligand-protein contact map. This network division allows contributions from ligand identity to be distinguished from effects of protein-ligand interactions on classification. We show, in agreement with recent literature, that dataset bias drives many of the promising results on virtual screening that have previously been reported. However, we also show that our neural network is capable of learning from protein structural information when, as in the case of binding mode prediction, an unbiased dataset is constructed. We develop a deep learning model for binding mode prediction that uses docking ranking as input in combination with docking structures. This strategy mirrors past consensus models and outperforms the baseline docking program in a variety of tests, including on cross-docking datasets that mimic real-world docking use cases. Furthermore, the magnitudes of network predictions serve as reliable measures of model confidence.

## I. INTRODUCTION

Computational techniques have long played a role in drug discovery efforts. These methods often involve training QSAR models that relate molecular features to targeted activities [1-3] or leveraging structure-based approaches that make use of the binding mode (pose), a three-dimensional description of the ligand interacting with the target protein. Protein-ligand docking programs, structure-based tools that are widely used by drug discovery modeling groups, [4-8] sample an ensemble of binding modes with the aim of optimizing a scoring function. Scoring functions are typically physics-based or physics-inspired, with emphasis on approximating trends in protein-ligand binding.

Techniques that combine neural networks with structure-based pose predictions have appeared in the literature [9-11]. Three-dimensional image-based convolutional neural networks [12-16] and, more recently, graph-based approaches [16-18] have been developed and applied to a variety of classification and regression tasks related to protein-ligand binding. When applied in conjunction with docking, the newer approaches take binding modes sampled by docking programs as input and 'rescore' them using a deep learning model. While initial results trained on popular 'benchmarking for docking' datasets such as DUD (Database of Useful (Docking) Decoys) [19] and DUD-E (Database of Useful (Docking) Decoys - Enhanced) [20] appeared promising, recently reported [21,22] comparisons with purely ligand-based informatics methods showed that purported virtual screening improvements were arising from ligand information alone. The fact that protein structural information has no demonstrable impact on classification points to bias in these datasets when used for deep learning.

We here present a simple, modular, graph-based deep learning architecture that allows for the ligand and protein-ligand contact information to be treated as inputs within a predictive model. This architecture is trained on two related tasks: (1) virtual screening (classifying ligands according to their activities) and (2) binding mode prediction (identifying a correct binding mode, given a protein-ligand pair). Since the effects of the ligand chemical structure and those of the protein-ligand binding model are readily separable within this framework, our model offers a clear paradigm for assessing the impact of these two feature types on performance. We show, in agreement with recent work [21,22], that for virtual screening tasks benchmarked on the DUD-E dataset, no significant improvement is observed when using protein-ligand complex structures as deep learning descriptors as opposed to purely two-dimensional ligand-based features.

The issue of bias in the virtual screening results motivates the identification of a task for which deep neural networks can successfully learn from the interactions present in protein-ligand complexes. We therefore develop a model for binding mode prediction, where ligand chemical structure alone cannot be predictive as, by construction, correct and incorrect binding modes are associated with each ligand. We show that our protein-ligand contact network can be trained on the PDBbind 2017 refined set [23,24] to yield a predictive model. While this model performs admirably according to the area under the receiver-operator characteristics curve (AUC, a metric that assesses how well models distinguish between classes), it does not show improvement over the baseline docking program in ranking the binding modes of individual target-ligand-pairs.

In order to show significant improvement over the baseline produced by docking results, we introduce an additional feature into our architecture: the docking pose rank. A model incorporating this feature with the above-described structural information is shown to improve performance over the baseline docking program on 5 of 6 independent test sets considered. This network serves as a consensus model between the representation learned from the three-dimensional complex and the pose-ranked output of docking. In particular, this model yields improvement for two cross-docking datasets taken from the literature [25,26]. In cross docking, a ligand is docked into an alternative target crystal structure that was not solved in co-complex with that specific ligand [27]; as such, cross docking is considered more difficult than the "self-docking" tasks the network is trained upon, since ligand-specific induced fit effects are not captured in cross-docking target structures.

Finally, estimating errors within machine learning models is an important, but often overlooked, objective[28]. The magnitude of the output classifier is shown to be useful in filtering out low-confidence poses based on an assessment of model precision.

This article is organized as follows. In Section II, we detail our method and parameter choices. In Section III, we present the results and insights from the model. Finally, in Section IV, we provide further discussion and conclusions.

## II. METHODS AND DETAILS

As in related deep learning-based approaches [11-13,18], docking programs are used to generate binding mode(s) for each target-ligand combination of interest. The binding modes are then taken as input into our deep neural network; as we will discuss later, additional inputs may also be considered. The output, i.e. the prediction of the model, can vary depending on the problem of interest. In this work, we develop networks to train on two different yet related binary classification tasks: compound activity against a target (virtual screening) or binding mode prediction. All deep learning models are written in Python using the TensorFlow library [29].

1. **Neural network architecture**

Chemical fingerprints have been used in cheminformatics and drug discovery for decades, and many approaches to defining fingerprints have been implemented [30-33]. One method in particular, ECFPX, generates fingerprints based on substructures using chemically bonded neighbors up to radius X/2 [32]. Neural-network-based approaches that utilize chemically bonded substructures have more recently been developed [34-36], and traditional fingerprint methods have been extended to include the protein-ligand contacts that characterize binding modes [37]. Integrating three-dimensional structures of protein-ligand complexes into a deep learning framework has been the subject of several recent papers. These methods include voxel-based CNN techniques [12,13,15] and methods that create embeddings of the local environment that ligand atoms experience [11]. More recently, graph-based deep neural networks have appeared in the literature [17,18]. We adopt a graph-based architecture that differs from those published previously in its minimalist design, particularly with regard to the property that no edges between protein nodes are considered. Furthermore, we utilize a modular form in which contributions from three-dimensional protein-ligand structure and intra-ligand chemical bonding can be readily separated for the purpose of assessing their respective contributions to the results. Fig. 1 schematizes our deep learning approach.

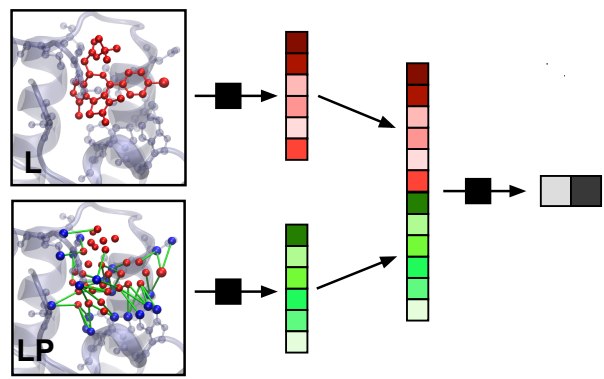

Figure 1: Schematic of the modular graph-based deep learning architecture. In the instance shown, two graphs - the ligand bonded graph ("L", red ligand nodes connected by red lines) and the ligand-protein contact map graph ("LP", red ligand and blue protein nodes connected by green lines) – are taken as input into a graph convolutional neural network. Internal representations (colored boxes) for each graph are concatenated and passed as input to fully connected neural network layers to make a classification prediction (represented by grayscale boxes). Structures are rendered with VMD [38].

Graph convolutional neural networks are of interest in a diverse set of fields [39] and have received attention for their natural fit in chemical problems to representing the bonded structure of a molecule. In the current application, the atoms of a molecule serve as nodes of a graph and edges are formed by the chemical bonds. The DeepChem implementation of graph convolutions (GraphConv) has been demonstrated to build a predictive regression model for binding affinity from complexes selected from the PDBbind database [36,40]. A high-level schematic of this graph convolution, as applied to a chemically bonded molecular structure, is shown in Figure 2A. We denote the representation vector resulting from the application of this graph convolution on the molecule's bonded structure as "L." This vector can be fed into additional layers to generate deeper representations and ultimately produce predictions. The details of our chosen graph convolutional architecture are adapted from DeepChem [36] and provided in more detail below.

We here extend the ligand graph convolution to capture the three-dimensional protein-ligand contact map that defines a given binding mode. Consider a graph where all protein atom and ligand atom sites are potential nodes. The determination of an edge between nodes is determined by a function $f(d_{ij})$ where $d_{ij}$ is the distance between protein atom, i, and ligand atom, j. In practice, we find that using a simple step function based on a protein-ligand site cutoff $r_c$ ,

$$f(d_{ij}) = \begin{cases} 1, & d_{ij} \leq r_c \\ 0, & d_{ij} > r_c \end{cases},$$

performs ably for the applications considered here, although more complex distance weighting schemes are possible. In order to simplify the model, ligand-ligand and protein-protein connections are disallowed, and nodes with no connections are omitted. As we find a relatively short cutoff works sufficiently well for the binding mode classification, only sites in or near the protein binding pocket need be considered (thus significantly decreasing the size of the input graph).

The above protein-ligand graph is then input into a graph convolutional network. We employ the same network architecture that is used for ligand graphs in DeepChem [36] and outlined below. This process yields what we refer to as the "LP" representation shown in Figure 2B. The LP vector can be fed into additional layers to produce a trainable model and output.

Our overall model is modular and can use the ligand network (L) or the protein-ligand network (LP) in isolation, or both simultaneously (L+LP); in the last case, the L and LP representations are concatenated and then fed into additional layers. By training and testing the L, LP and L+LP models on the same data, the contributions of the ligand chemical structure and the protein-ligand binding mode to the combined model can be readily evaluated. Additional features beyond these two graphs can be added within this framework. For example, we later show that it is advantageous in binding mode prediction to include the rank of a pose predicted by the docking program chosen to generate our input structures (R; Fig 2C). Fig. 2D summarizes how the internal representations generated in Figs. 2A-C can be combined and fed through additional neural network layers in order to generate a trainable model and a predictive output. The tasks considered here - virtual screening and binding mode prediction - are formulated as binary classification problems, and thus rely on a final softmax layer with output dimension 2 to estimate class probabilities.

2. **Choice of docking program and datasets**

As mentioned above, docking programs sample an ensemble of binding modes with the aim of optimizing a scoring function. These scoring functions are drawn from sources on a continuum ranging from physics-based to empirical and knowledge-based to machine learning [41]. Assessing the performance and improving the quality of docking programs is a longstanding goal of the field [42-47]. To generate binding modes for use as input to our model, docking simulations were carried out with AutoDock Vina [8] on each protein-ligand pair of interest. AutoDock Vina is widely-used docking program whose scoring function has been optimized on the PDBBind "core" dataset [23,24]. Many other docking packages are available [5-7], and our method can be readily trained with the docking output of any such program. Here, docking was performed using a 27 Å cubic search box centered on the position of the reference ligand conformation with an exhaustiveness parameter of 16, matching a protocol employed in the literature [11,48]. Higher values of the exhaustiveness parameter were tested with no difference in overall performance observed. For the virtual screening task, only the top ranked pose produced by AutoDock Vina is considered for each complex. For binding mode prediction, more than one pose is required by construction, and the top 20 ranked poses output by AutoDock Vina are analyzed.

An output prediction of activity defines the virtual screening task. To illustrate performance, we test and train the L, LP, and L+LP networks on the DUD-E dataset [20]. The DUD-E dataset, a revision of the earlier DUD dataset [19], contains collections of active and decoy ligands for 102 protein targets. Both DUD and DUD-E have previously been used to compare the performance of different docking programs and to train deep learning networks for activity prediction [11,14]. Ligands are labeled as positive or negative according to their presence in the active or decoy (presumed inactive) set, respectively, and decoys have been chosen for each individual target to have similar molecular weights and other properties to the known actives. A three-fold cross validation is used as in Ragoza, et al. [14]; but in distinction to that work, our splitting is done according to ligand dissimilarity rather than target sequence similarity (see Supporting Information).

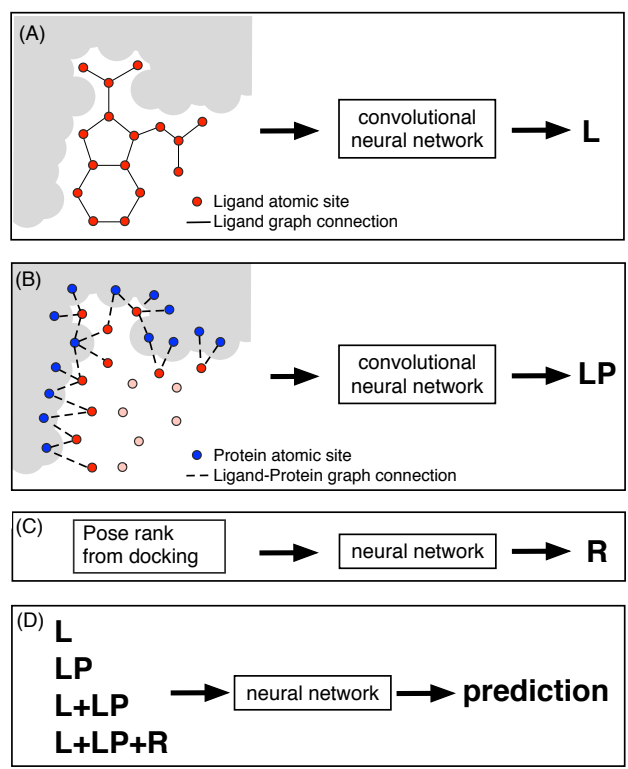

Figure 2: Two graphs are used to characterize the protein-ligand complex. In the "L" graph (Fig. 2A) ligand sites (red circles) are connected according to the presence of chemical bonds (solid lines). In the "LP" graph (Fig. 2B), protein sites (blue circles) in contact with ligand sites form a separate network. The protein surface is shown in gray. Each graph is processed by a convolutional neural network. The resulting internal representations, individually (L, LP) or combined (L+LP), are fed into further layers to yield a predictive output (Fig. 2D). Additional features, such as the pose ranking from docking programs, can be added as input to the network (Fig. 2C) and then combined with the other elements in a modular fashion (L+LP+R).

For the binding mode prediction task, we train the L, LP, L+LP, and L+LP+R networks on the refined subset of the PDBbind 2017 database. Seventy (70) percent of the data are used for training and validation and the remaining 30% serves as the test set. In order to roughly mimic a procedure of training on past data to predict future data, we use a test/training split based on alphanumerical order according to each complex's PDB ID code. Complexes are labeled as correct if they fall within 2.0 Angstroms heavy-atom RMSD of the reference ligand structure, a standard cutoff in the field (see e.g. Ref. 44). Correct poses are labeled "positive," and those that fall outside the cutoff are labeled as "negative."

3. **Network details and hyperparameters**

Atom sites in our model are defined by the atomic descriptors generated using the RDKit cheminformatics library [49]. Only heavy atoms are explicitly treated within the graphs. Adapted from the DeepChem graph node representation, the atomic features are: element name, hybridization, number of attached hydrogens, formal charge, absence or presence of an unpaired electron, and whether or not the atom participates in an aromatic substructure. Input representation vectors are generated via a concatenation of one-hot encoded vectors over all atomic features.

The graph topology is represented as a connectivity list that specifies the nearest neighbors of each node. The graph convolutional network architecture adapted from DeepChem is composed of alternating nearest-neighbor convolution and pooling layers. Convolution and pooling units can be repeated an arbitrary number of times, but a depth of 3 was deemed sufficient for many tasks in the original implementation [36]. A single dense neural layer follows the final pooling layer in the network; the dense node dimension is reduced using a gather operation that takes both the mean and maximum across nodes. A more detailed discussion of this architecture can found in Altae-Tran, et al. [36]. In our modular approach, this representation can be combined with the results from other inputs (graphical or other) and fed to further dense layers, ultimately yielding an output prediction. A schematic of the network is given in Figure S1.

Initial hyperparameters for the "L" network were adopted from DeepChem and were also used as an initial hyperparameter state for the LP model. L2 regularization was applied with a scaling coefficient of 0.0005, as determined by exploration of several parameter choices.

For the "L+LP+R" network, select hyperparameters of the network (see SI) are optimized more systematically using random exploration of hyperparameter space [50]. In this procedure, eight validation sets were generated from random 10%/90% splits of the training set; independent models were trained on the 90% partition of each splitting. This training program was repeated for 200 sets of parameters. Hyperparameters were chosen to maximize the fraction of correct top ranked binding modes when averaged over the eight (10%) validation sets. Hyperparameter values are given in Table S1. Results reported for L+LP+R on the PDBbind test set (Section III-2) are averaged over output from the eight trained models corresponding to the chosen hyperparameter set. Results reported for L+LP+R in Section III-3 are averaged over five training instances using the full PDBbind refined set.

## III. RESULTS

### 1. Role of protein structure in virtual screening and binding mode prediction tasks

Table 1 reports the area under the receiver-operator curve (AUC) observed for the ligand-only (L) [36], ligand-protein (LP), and combined dual graph L+LP approaches measured against baseline AutoDock Vina results for virtual screening (training and testing on DUD-E) and binding mode prediction (training and testing on PDBbind refined). The AUC represents the integral of the true positive rate as a function of the false positive rate for an ordered list of model predictions. AUCs are particularly useful for evaluating performance on the unbalanced datasets that one typically encounters in drug discovery classification tasks, where active ligands are often in the minority.

The pitfalls of using DUD-E to evaluate structure-based deep learning models can been seen upon inspection of the AUCs given in the first column of Table 1. The top-ranked poses produced by AutoDock Vina yield an AUC of 0.70, which is in fair agreement with previously reported values[14]. Three of the four machine learning methods show a significant improvement over Vina, in agreement with the results of other deep learning approaches in the literature [11,14]. However, as we show below and as seen in very recent work [21,22], the observed improvement is not due to learning from 3D structural information.

Because of the modular nature of our network, we can use the L and LP networks independently to test the relative contributions of ligand identity alone (through internal bonded structure) and the contacts between ligand and protein sites that define the binding mode. For virtual screening tasks trained and tested on the DUD-E dataset, the L network significantly outperforms the LP network. Therefore, it is shown that ligand-identity is driving the performance of the combined (L+LP) model. Furthermore, standard cheminformatics representations such as the Morgan circular fingerprint (analogous to ECFP4) [32] available in the popular package RDKIT [49] fed into a random forest classifier produce similar results to the "L" network, and far surpass the result where the ligand-protein graph serves as the sole input. Thus, while virtual screening/DUD-E benchmarks are intended to demonstrate that deep learning methods learn from the 3D structures of protein-ligand complexes, those models are actually learning from ligand-only information and produce equivalent results to traditional cheminformatics approaches.

As noted in recent work [21,22] this result is likely due to some bias in choosing active and decoy molecules in the DUD-E dataset that can be readily "solved" by machine learning models. In other words, machine learning models can identify active and decoy molecules in the DUD-E datasets independent of their protein interactions. Similar issues related to applying machine learning techniques to such datasets have appeared in the literature [51,52]. This finding is also apparent upon close examination of the hyperparameter optimization of the DeepVS model of Pereira, et al. which shows that even if protein structural information is left out of the feature set, the performance on the DUD dataset worsens only slightly [11].

We next consider the task of binding mode prediction in the context of training and testing on structures in the PDBbind refined dataset [23,24] (second column of Table 1). These results show that the choice of dataset and/or task, and not a failure in network architecture, limits learning from protein-ligand interactions in DUD-E/virtual screening. In the case of binding mode prediction, the labels are related to the "3D" RMSDs of binding modes rather than the "1D" activities of protein-ligand combinations, and so the L+P network is encouraged to learn from protein ligand contacts rather than ligand identities. Indeed, the "L" network (which only takes chemical structures of the ligand as input) cannot distinguish the three-dimensional orientations that define binding modes. In this case, it can be seen (Table 1, second column) that the LP network significantly outperforms the L network. The "L" network yields a result close to random (AUC: 0.54). Though the L network might still be gleaning some information from the general "dockability" of ligands (which may relate to size or other properties), the LP network (AUC: 0.83) clearly captures the fundamentals of binding mode prediction that the L network cannot. The combined L+LP network primarily reflects the LP results, although it reaches slightly higher values (AUC: 0.86).

Table 1 also shows that the binding mode prediction AUC reported for AutoDock Vina is significantly lower than the AUC achieved by our model. However, this result is not indicative of real-world performance, which is usually measured according to the ranking of binding modes for a single target-ligand pair. Instead, these data show that the probabilities output by our model are, as absolute numbers, more predictive of pose correctness than docking-generated scores or "binding energies" across different sets of targets and ligands. Thus, the output probability can be thought to provide a measure of confidence for a given binding mode. We will return to this point below.

**Table 1: Comparative performance (as measured by AUC) of methods trained and applied on virtual screening and binding mode prediction tasks.**

|  | Virtual Screening – DUD-E | Binding Mode Prediction – PDBbind |
|---|---|---|
| AutoDock Vina | 0.70 | 0.66 |
| Morgan Radius 2 / RF | 0.83 | --- |
| L | 0.84 | 0.54 |
| LP | 0.64 | 0.83 |
| L+LP | 0.82 | 0.86 |

## 2. Improving binding mode prediction from docking using a novel architecture

Having demonstrated that the 3D structures of protein-ligand binding poses can be used to train a binding mode prediction model, we next tested if such models could improve upon the results generated by the initial docking program.

In Table 2, we report, in addition to the AUC, the percentage of top ranked (ranked "1") binding modes that are correct. The condition of correct is met when a binding mode falls within 2.0 Å RMSD of the heavy-atom reference ligand coordinates. It can be seen that in terms of this metric, which is often of most interest to the end user, Vina outperforms the L+LP model's 're-scoring' of its binding modes. This result is generally in agreement with published results for a voxel-based deep learning model [13]. This single ligand-protein pair test yields results more favorable to the docking program than the AUC measure, as the AUC indicates how well docking scores/energies rank all poses in the test set (see Table 2).

We next introduce the subnetwork schematized in Figure 2C to include pose ranking as an additional feature of our model. The output of a simple pose rank-fed sub-network is combined with the internal representations generated by the L and LP sub-networks (Figures 2A and 2B); this combined representation is directed through a final set of three dense layers to yield a prediction that describes a "consensus" of the inputs. Consensus models have been developed that use multiple docking programs to generate an improved result [53,54]. Here, we use deep learning to combine protein-ligand structural features with the docking rank to produce a new prediction. We label this model "L+LP+R." Hyperparameters for this system were optimized as discussed in Section II-3.

**Table 2: Comparison of model performance on the independent test set drawn from the PDBbind database, as measured by AUC and the fraction of correct top ranked binding modes.**

|  | AUC | top ranked fraction correct |
|---|---|---|
| Autodock Vina | 0.66 | 0.364 |
| L+LP | 0.86 | 0.306 (0.003) |
| L+LP+R | 0.90 | 0.380 (0.004) |

The results of the "L+LP+R" model on our PDBbind test set are given in Table 2 and Figure 3. It can be seen that the "L+LP+R" model is the top performer, significantly improving on the "L+LP" model and improving the Vina result by roughly 5 percent with respect to the fraction of top ranked poses that are correct (Table 2). Figure 3 shows the cumulative fraction of systems that contain at least one correct pose up to a given pose number. For example, at x=5, the y-values indicate the fraction of systems that have at least one "positive" pose in the top 5 according to the rankings specified by each model. It can be seen that the "L+LP+R" model maintains roughly the advantage seen on the first pose until approximately x=10, where all plots in Figure 3 start to level off. In 65% of systems in our test set, the docking program samples a correct mode in at least one of the 20 rank positions. As our model only re-ranks docking output, improvement in sampling error [44] is beyond the scope of our method. Results renormalized by this sampling error are included in Table S2.

A very recent publication [18] claimed to achieve a 5-7% improvement over baseline docking results on the PDBbind 2018 dataset, seemingly outperforming our L+LP model and roughly matching the results of our L+LP+R model. However, unlike in the present results, the training and testing datasets in Ref. 18 omit borderline poses with RMSDs between 2 Å and 4 Å. In other words, positive poses in that study have RMSDs less than 2 Å and negative poses have RMSDs greater than 4 Å. Marginal poses are presumably some of the most important yet difficult to classify in real world applications. If these poses are filtered from our dataset, we find that the L+LP model exhibits increases in AUC and the fraction top ranked metric and shows an ~12% relative improvement over our docking baseline (Tables S5 and S6 in the Supporting Information), results comparable to those presented in Ref. 18. The results produced by the L+LP+R model using the filtered dataset are also boosted (roughly 30% above our docking baseline). However, we feel these seemingly positive results do not adequately reflect model performance in production environments, as exclusion of valid but difficult-to-classify data points is not recommended for building robust machine learning models.

The values of the probabilities output by our model are also good measures of the model's confidence in those predictions. This fact can be seen by evaluating the positive predictive value (precision) of the model over various thresholds (see Figure 4, left panel). Further details of how this plot is generated is given in the Supporting Information. Using a threshold of 0.9 the "L+LP+R" model has a precision of approximately 90%. Negative pose prediction is shown in Figure 4, right panel. As the dataset is unbalanced, the overall higher negative predictive value is expected given our AUC measurements.

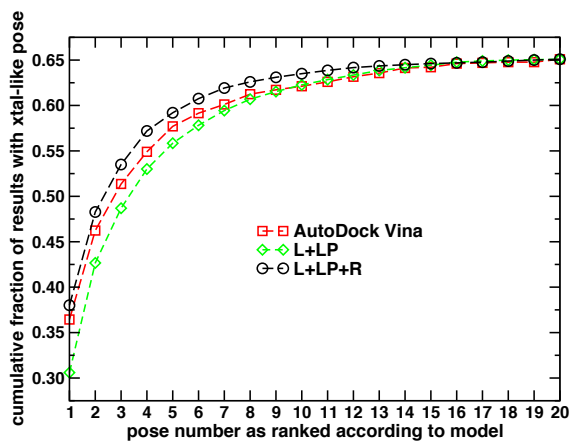

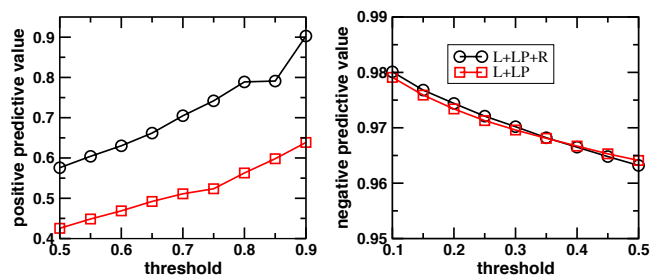

Figure 3: Cumulative fraction of results with crystal-like (within 2 Å RMSD) binding mode plotted versus pose number as ranked according to AutoDock Vina (red squares) the L+LP model (green diamonds), and the L+LP+R model (black circles).

## 3. Additional test sets and cross docking results

We next tested our model on a set of small sets outside of PDBbind. For this test, we retrained our models on the whole of the PDBbind refined test set using the already optimized hyperparameters (see Table S1). Sets were constructed from a selection of targets in PDB database that correspond to bromodomain 4 (BRD4), kinases, as well as the PDBs from which the targets of the DUD-E dataset [20] are drawn. Complexes that were repeated from the PDBbind refined set were removed from these test sets. Details of the datasets are given in the Supporting Information.

Table 3 shows the performance of the "L+LP+R" model compared against AutoDock Vina using the fraction of correct top ranked modes metric. Performance on the BRD4 and DUD-E test sets is in line with the performance boost seen in the PDBbind refined test set. Autodock Vina clearly out-performs "L+LP+R" in the Kinase test set. As in any machine learning system, our model can be improved systematically by the introduction of more training data tailored to a problem at hand, so more kinase training data might be used to boost performance in the future. Results renormalized by their respective sampling error are included in Table S3.

A task of particular interest is cross docking, which was defined earlier. Cross docking is a more difficult problem than self-docking, and the approach is more realistic with respect to how docking would be used within a drug discovery workflow. We tested our method on subsets of the ASTEX non-native set [25] and CSAR 2010 set [26], which both contain curated cross docking examples from the PDB. For the ASTEX set, we used a subset of structures that we were able to successfully dock and featurize within our trained model, ignoring, for example, structures with atomic properties not present in our training set. In the case of CSAR, well-aligned structures (i.e. structures for which the binding pockets of the various target proteins corresponding to a given ligand are aligned) from the PDBbind 2017 refined set were selected. Note that unlike the results presented in Table 3, PDBbind structures are kept in the self-docking companion sets for completion. The list of structures and ligands comprising the datasets are given in the supplement.

Figure 4: (Left panel) positive predictive value (precision) of the L+LP+R model (black circles) and L+LP model (red squares) as a function of the threshold for positive / negative labeling. (Right panel) In parallel to the left panel, the measured negative predictive value vs. threshold for both models.

In Table 4, we show our baseline Vina results and the results of the "L+LP+R" model. The "L+LP+R" model exhibits improvements relative to Vina of roughly 10-15% for both cross docking sets. To our knowledge, the present method is the first test of deep learning enhanced docking approaches on cross docking datasets, and the results are encouraging for the utility of our deep learning model in drug discovery use cases. Both Vina and "L+LP+R" score highly on the self-docking companion sets, although Vina slightly outperforms the "L+LP+R" model on the ASTEX-self set. Results renormalized by their respective sampling errors are given in Table S4.

Table 3: Performance of L+LP+R model against AutoDock Vina for a set of small test sets.

| Test set | Number of complexes | Vina top ranked fraction correct | L+LP+R top ranked fraction correct (ERR) | L+LP+R AUC |
|---|---|---|---|---|
| BRD4 | 179 | 0.315 | 0.342 (0.01) | 0.84 |
| DUD-E | 84 | 0.422 | 0.455 (0.007) | 0.93 |
| Kinase | 211 | 0.502 | 0.469 (0.005) | 0.88 |

**Table 4:** The L+LP+R model tested against AutoDock Vina results for two cross-, and their corresponding self-, docking datasets.

| Test set | Number of complexes | Vina top ranked fraction correct | L+LP+R top ranked fraction correct (ERR) | L+LP+R AUC |
|---|---|---|---|---|
| ASTEX-SELF | 61 | 0.754 | 0.744 (0.018) | 0.94 |
| ASTEX-CROSS | 1055 | 0.340 | 0.377 (0.003) | 0.90 |
| CSAR-SELF | 89 | 0.652 | 0.676 (0.011) | 0.95 |
| CSAR-CROSS | 262 | 0.359 | 0.423 (0.004) | 0.93 |

**DISCUSSION AND CONCLUSIONS**

Over the past several years, applications of deep learning methods to drug discovery workflows have been explored with increasing frequency. One area of particular interest regards combining deep learning approaches with the output of structure-based docking models. While early reports were encouraging [11,12,14], more recent work has raised concerns about spurious results associated with dataset bias [21,22]. Therefore, it was hitherto unclear if deep learning models were actually capable of learning from 3D structures in a productive fashion.

In this work, we have developed a new set of deep learning models that take the output of docking programs as input. Our models can be fashioned to classify activity (virtual screening) or the accuracy of binding modes (binding mode prediction), and our underlying networks are based on a modular graph convolutional architecture that can take multiple graphs as input. Specifically, distinct graphs are used to represent the ligand bonded structure and the protein-ligand contact map as a binary (in/not in contact) system. Additional features, such as the docking pose rank, can be added to the network as well.

We find that previously reported improvements in docking-based virtual screening that were credited to the deep learning on the protein co-complex instead result from learning on ligand bonded topologies alone, in agreement with other recent work [21,22]. However, we show that our deep learning model can indeed learn from 3D protein-ligand structures in binding mode prediction tasks, giving further evidence that dataset bias is the culprit in past specious results.

We also find that, by combining our deep learning model with the initial pose rank output from docking, we are capable of producing a consensus model that improves upon the results of our baseline docking program. This improvement is demonstrated on the initial test set as well as several smaller datasets. In particular, we show the method's utility in cross docking, a more realistic use-case with respect to pharma-based applications than the oft-studied self-docking.

A very recently published work [18] also combines docking-generated structures with a graph-based neural network and applies that model to the tasks of virtual screening and binding mode prediction. While the qualitative results – and apparent pitfalls – with respect to virtual screening presented in this past work are broadly in agreement with what is known from the literature [14,21,22], their results on binding mode prediction require some comparison with the present work. Specifically, their results show improvement over the baseline docking using training and test sets that omit "borderline negative" poses (between 2 Å and 4 Å RMSD). We are able to reproduce similar improvements using this methodology (see Supporting information). However, our primary results focus on unfiltered data sets that better reflect real-world use cases in which reference structures are unknown, and thus where determination of RMSDs and selective elimination of structures is not possible.

Our results provide evidence that deep learning can be a useful tool in structure-based drug discovery. Since one cannot hope to understand protein-ligand interactions without first identifying the correct ligand binding pose, binding mode prediction represents a foundational tool that can feed into other drug discovery workflows. In addition to improving virtual screening with protein structure and optimizing small molecule leads with higher-level free energy calculations, pose prediction could also become an important aspect of deep learning-based generative modeling methods that incorporate structural information from a desired binding pocket.

Furthermore, the graph-based method presented here offers an attractive alternative to voxel-based approaches that have been applied by others to three-dimensional structural inputs, as the translational and rotational invariance of the graph representation facilitate featurization and avoid the concerns over global alignment seen with voxels. Indeed, pair distance-based energy functions have been a cornerstone of physics-based molecular simulation for decades; graphs provide a similar level of simplicity and modularity in featurizing physical systems for input into deep neural networks. While the graphs used in the present work are simple binary contact maps, the method can be extended to include distance-weighted edges or any number of alternative node- or edge-based features. Such a flexible framework for translating between physical and informatic data structures should prove useful as combinations of physics-based and deep learning models become more prevalent. In this case, only the marriage of physics-inspired and informatic techniques (here, a docking scoring function and a machine learning model trained on pose coordinates, respectively) proved capable of improving on the model used as our baseline. Going forward, similar synergies between features derived from physical laws that govern molecular structure and dynamics and data generated by observation are likely present in a wide array of systems. Even if ever larger datasets shift the balance toward informatics in the future, the complementarity of physics and data certainly warrants further investigation.

**Corresponding Author**

* email: jamorron@us.ibm.com

**ACKNOWLEDGMENT**

We would like to thank Cicero Nogueira dos Santos for many useful discussions.

# Supporting information for "Combining docking pose rank and structure with deep learning improves protein-ligand binding mode prediction"

Joseph A. Morrone*, Jeffrey K. Weber, Tien Huynh, Heng Luo, Wendy D. Cornell

Healthcare & Life Sciences Research, IBM TJ Watson Research Center
1101 Kitchawan Road, Yorktown Heights, NY 10598, USA

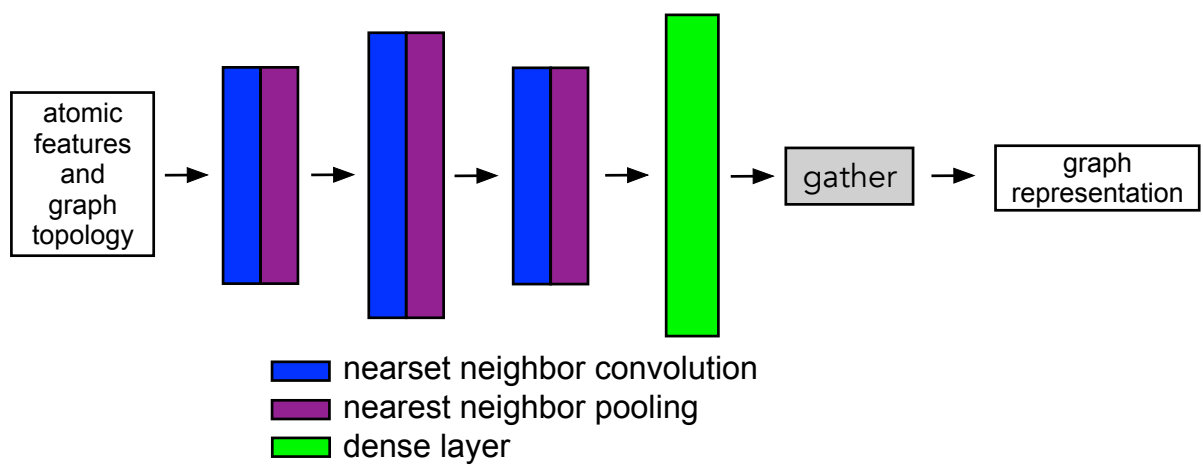

Figure S1: Schematic for the graph convolutional neural network used in this study. This element of our network is adapted from DeepChem. The graph is composed of molecular connectivity (topology) and the atomic features that define the nodes. Features are passed into a set of alternating nearest neighbor convolution and pooling layers. A depth of 3 is used here and in the original implementation. After a single dense layer, a gather operation combines the output from all previous nodes.

**Table S1 Hyperparameters for L+LP+R model**

|            | Nconv 1 | Nconv 2 | Nconv 3 | N dense | N final 1 | N final 2 | L2     |
|------------|---------|---------|---------|---------|-----------|-----------|--------|
| L+LP       | 64      | 128     | 64      | 128     | ---       | ---       | 0.0005 |
| L+LP+R opt.| 256     | 32      | 192     | 336     | 208       | 288       | 0.0003 |

# DEFINITION OF AUC

The AUC computation involves counting the number of negative entries (inactive or decoy ligands) that are ranked in the top N+ entries where N+ is the number of positives (active ligands). The formula is given as:

$$\text{AUC} = 1 - \frac{1}{N_+}\sum_{i=1}^{N_+}\frac{N^i_-}{N_-}.$$

# DEBIASED DUD-E SPLITTING

Our goal in generating debiased train/test splits for DUD-E involves splitting the target sets into test and training sets in a way that minimizes the difference between active and decoy (presumed inactive) ligands across target sets. The splitting uses dataset bias ideas similar to those expressed in Ref. 52 in the main text.

1. Ligands are represented by an ECFP4-like fingerprint using RDKit (Ref. 49 in main text).
2. Ligands are separated into groups according to their associated target.
3. The distribution of Tanimoto similarities is computed between ligands in target sets X and Y, and then also split according to their labels (A=active, I=decoy/presumed inactive): $\rho^{XY}_{AA}$, $\rho^{XY}_{AI}$, $\rho^{XY}_{IA}$, $\rho^{XY}_{II}$.
4. The bias between two target sets is sensitive to the difference between the respective active and inactive distributions and is given by the following equation:

$$bias^{XY} = \text{KL}(\rho^{XY}_{AA}, \rho^{XY}_{AI}) + \text{KL}(\rho^{XY}_{AI}, \rho^{XY}_{AA}) + \text{KL}(\rho^{XY}_{II}, \rho^{XY}_{IA}) + \text{KL}(\rho^{XY}_{IA}, \rho^{XY}_{II})$$

where KL(a,b) is the Kullback-Leibler divergence between distributions a and b.

$$\text{KL}(a,b) = \sum_i a_i \ln\left(\frac{a_i}{b_i}\right)$$

5. The splitting is chosen to minimize the global bias (the sum of the target-pairwise biases) using a Monte Carlo procedure.

# DEFINITION OF POSITIVE PREDICTIVE VALUE AND NEGATIVE PREDICTIVE VALUE

The positive predictive value (PPV) or precision and negative predictive value are defined as:

$$\text{PPV} = \frac{N_{TP}}{N_{TP} + N_{FP}}$$

$$\text{NPV} = \frac{N_{TN}}{N_{TN} + N_{FN}}$$

where $N_{TP}, N_{FP}, N_{TN},$ and $N_{FN}$ are the number of true positive, false positive, true negative and false negatives, respectively. The boundary between positively and negatively classified samples is set by the threshold which is varied on the x-axis of Fig. 4 in the main text.

# FRACTIONS RE-NORMALIZED BY SAMPLING ERROR

**Table S2: Fraction of top ranked renormalized by sampling error. Compare with Table 2 in the main text**

|  | Fraction top ranked correct |
|---|---|
| Autodock Vina | 0.559 |
| L+LP | 0.470 |
| L+LP+R | 0.584 |

**Table S3: Fraction of top ranked renormalized by sampling error. Compare with Table 3 in the main text.**

| Test set | Number of complexes | Vina top ranked fraction correct | L+LP+R top ranked fraction correct | Fraction sampled |
|---|---|---|---|---|
| BRD4 | 179 | 0.392 | 0.426 | 0.803 |
| DUD-E | 84 | 0.778 | 0.839 | 0.542 |
| Kinase | 211 | 0.602 | 0.562 | 0.834 |

**Table S4: Fraction of top ranked renormalized by sampling error. Compare with Table 4 in the main text.**

| Test set | Number of complexes | Vina top ranked fraction correct | L+LP+R top ranked fraction correct | Fraction sampled |
|---|---|---|---|---|
| ASTEX-SELF | 61 | 0.904 | 0.892 | 0.8341 |
| ASTEX-CROSS | 1055 | 0.537 | 0.595 | 0.6332 |
| CSAR-SELF | 89 | 0.795 | 0.824 | 0.8202 |
| CSAR-CROSS | 262 | 0.560 | 0.660 | 0.6412 |

# BINDING MODE PREDICTION RESULTS ON "FILTERED" PDBBIND DATASET

A very recent work (Ref. 18 in the manuscript) has published results on a graph-based deep learning model for binding mode prediction. This work uses datasets that classify poses as either (1) less than 2 Å RMSD from the reference structure or (2) greater than 4 Å from the reference. Prospective binding modes between 2 Å and 4 Å are omitted from the datasets for both training and testing. Their results showed improvements over the baseline docking of 5-7%.

Table S5 shows results derived from our networks using this same data set filtering technique. We see similar magnitudes of improvement over our baseline docking results (reported in Table 2 in the main text) when compared with Ref. 18. In Table S6, we show that similar performance is also achieved after using the trained models reported in the main text (that is, models trained on all available poses) but removing poses between 2 Å and 4 Å RMSD from the test set.

Finally, Table S7 shows the results of training on filtered poses but testing on unfiltered poses. As can be seen upon comparison of Tables S5-S7, the observed improvements appear to result from filtering borderline poses from the test set.

**Table S5: Trained and tested on filtered poses generated from PDBbind dataset.**

| Model | AUC | Top-Fraction |
|---|---|---|
| L+LP | 0.89 | 0.407 |
| L+LP+R | 0.92 | 0.471 |

**Table S6: Trained on all available poses, tested on filtered poses generated from PDBbind dataset.**

| Model | AUC | Top-Fraction |
|---|---|---|
| L+LP | 0.89 | 0.394 |
| L+LP+R | 0.91 | 0.463 |

**Table S7: Trained on filtered poses, tested on unfiltered poses generated from PDBbind dataset.**

| Model | AUC | Top-Fraction |
|---|---|---|
| L+LP | 0.86 | 0.313 |
| L+LP+R | 0.90 | 0.387 |